# Optical Curtain Effect: Extraordinary Optical Transmission Enhanced by Antireflection

Yanxia Cui[1,2,*], Jun Xu[3], Yinyue Lin[1], Guohui Li[1], Yuying Hao[1], Sailing He[2], and Nicholas X. Fang[3]

[1]Department of Physics and Optoelectronics, Taiyuan University of Technology, Taiyuan 030024, China

[2]Centre for Optical and Electromagnetic Research, State Key Laboratory of Modern Optical Instrumentation; JORCEP (KTH-ZJU Joint Center of Photonics), Zhejiang University, Hangzhou 310058, China

[3]Department of Mechanical, Massachusetts Institute of Technology, Cambridge, Massachusetts 02139, USA

Corresponding author: yanxiacui@gmail.com

## Abstract

In this paper, we employ an antireflective coating which comprises of inverted π shaped metallic grooves to manipulate the behaviour of a TM-polarized plane wave transmitted through a periodic nanoslit array. At normal incidence, such scheme can not only retain the optical curtain effect in the output region, but also generate the extraordinary transmission of light through the nanoslits with the total transmission efficiency as high as 90%. Besides, we show that the spatially invariant field distribution in the output region as well as the field distribution of resonant modes around the inverted π shaped grooves can be reproduced immaculately when the system is excited by an array of point sources beneath the inverted π shaped grooves. In further, we investigate the influence of center-groove and side-corners of the inverted π shaped grooves on suppressing the reflection of light, respectively. Based on our work, it shows promising potential in applications of enhancing the extraction efficiency as well as controlling the beaming pattern of light emitting diodes.

Since the discovery of extraordinary optical transmission (EOT) in 1998 [1], metallic nano-structures have attracted significant attention because of their novel ability to excite propagating surface plasmon polariton (SPP) or localized surface plasmon modes (LSP) [2-4]. Usually, exotic phenomena, e.g., the well-known EOT [1, 5-7], Fano resonance [8, 9], hot spot effect [10, 11], etc., accompany with the desired metallic nano-structures. In advantage of state-of-the-art nanofabrication techniques, the metallic nano-structures are playing remarkable roles in current nano-science and have been used in many aspects including extracting more light from light emitting diodes [12, 13], harvesting solar energy [14, 15], improving sensitivity of biosensors [16, 17], focusing light with sub-diffraction limit resolution [18-20] and so on.

One simple sample of the metallic nano-structures is a single metallic nanoslit which can diffract light into all radial directions in cylindrical wave manner [21]. The behaviour is equivalent to a line current source, i.e., a point source in the wave propagating plane. Particularly, a non-resonant metallic nanoslit performs negligible influence on the incident field at the input opening. It can read out the signal of incident field (both amplitude and phase) at its input and accordingly transfer that signal into the output region [22]. If the nanoslit is performing on resonance, one can obtain a greatly enhanced transmission of light passing through the slit in comparison with the integrated incident energy only on the slit opening (i.e., EOT). EOT can be more efficiently produced by dressing the input surface of the metallic slit with some grooves or other resonant cavities [5, 23-25]. By corrugating the output surface of the slit, the uniform diffraction pattern in the output region can be tuned into some focusing or deflection profiles [23, 26] as well. Additionally, a group of metallic nanoslits can strongly interact with each other to greatly enhance the transmission efficiency defined as the ratio of the integrated energy flow over the output aperture to that over the input aperture (i.e., the slit opening) [27, 28]. Studies on shaping the metallic nanoslits are on-going [29, 30]. Some complex structures, such as double layered metallic nanoslit array have also been investigated [31, 32] to further explore the EOT phenomenon. Inside the laboratory, researchers have already introduced metallic nanoslits into the technique of nanolithography in both near field [33] and far field [34].

In our previous work [35], it is reported that light coming out of a metallic nanoslit array can generate a unique optical curtain effect, i.e., the formation of a spatially invariant field distribution in the output region. Based on the physical model in the wave-vector domain, it is known that the optical curtain effect is actually the superposition of two diffracting plane waves (±1 orders). At the input openings of the nanoslits, both amplitude and phase of light have been necessarily manipulated by a certain metallic coating. However, the coating film composed of metallic nanostrips highly reflects the incident light which results in a very low transmission. In this paper, in order to suppress the reflection by the metallic strips, we carve some corrugations on the strips to form some inverted $\pi$ shape grooves, an array of which is then used to manipulate the light incident on the input openings of a periodical nanoslit array. As a result, not only the optical curtain effect in the output region is retained, but also the transmission efficiency of light through the metallic nanoslit array at normal incidence can reach 90% approximately, which might have promising applications in light emitting diode design and photolithography.

Fig. 1 shows the structure diagram of the proposed periodic metallic system. The inverted $\pi$ shaped groove is formed by two steps: first, etching a center-groove (of width $B_1$ and height $B_2$) into the original metallic nanostrip (of width $W_p$ and height $h$); second, etch two additional side-corners (of width $B_3$ and height $B_4$) into the nanostrip. Later, we will show that the center-groove and the side-corners can work as anti-reflection elements separately. The periodicity of the inverted $\pi$ shaped groove is $P$. A plane wave of TM polarization (magnetic field $H_y$ perpendicular to the x-z plane) is incident normally (along -z direction) on them. A metallic film with nanoslits (of width $W_s$, height $t$, and periodicity $P/2$) is placed under the inverted $\pi$ shaped grooves with a distance $d$. These nanoslits under illumination work as a series of line current sources, which deliver cylindrical waves from their openings. Geometrical parameters of the nanoslits are fixed with $W_s$ = 55 nm and $t$ = 65 nm, the periodicity is fixed at $P$ = 1100 nm and the separation $d$ is equal to 140 nm. We note that the gap between neighbouring grooves indicated as $P$-$W_p$ should be in a reasonable range, since large gap will cause some oscillated fringes along $z$ direction, but small gap will block too much incident light. In our design, we make a compromise of gap size by setting the nanostrip width $W_p$ = 740 nm.

Without losing the generality of noble metal, the metal structures are made of silver with its permittivity obtained from Ref. [36]. Except for the spectra study in Fig. 3, we fix the incident wavelength at $\lambda_0$ = 1 $\mu$m and the corresponding relative permittivity of silver is $\varepsilon_m$ = -45.7+2.84$i$. The metallic nano-system is free-standing in air for convenience of study, while in practice a substrate is required and the structural parameters should be tuned accordingly to keep the optical curtain distribution. All the results are calculated using a 2D finite element method (Comsol, RF model, TM mode) [37]. Periodical boundary condition is applied along $x$-axis and perfect matched layers are applied along $z$ direction to avoid artificial reflections from the boundaries. The validity of our numerical method has been confirmed by comparing the results with those calculated with Rigorous Coupled-wave Analysis (RCWA) method [38].

First of all, we check the system with un-etched metallic nanostrips (i.e., with flat top surface) to illustrate that the transmission cannot be quite high by tuning the strip height. The inset of Fig. 2a shows the relation of the transmission efficiency ($\eta$) and the strip height ($h$) when $h$ is tuned from 20 to 1500 nm. It is seen that the transmission of light through the nanoslit array suffers a very small fluctuation scope (from 13.7% to 20.1%) which is generated by the effect of Fabry-Perot resonance in the air gap between neighbouring nanostrips. The largest transmission is obtained when $h$ = 440 nm. The corresponding field distribution ($|H_y|$) is displayed in Fig. 2a. In the figure, we observe that there is no field variation along $z$ direction in the output region. In addition, there are two peaks and two valleys in each period along the $x$ direction, which we named it as optical curtain effect [35]. Here, we emphasize that reflection of light from the flat top surface of the metallic strip is quite strong in Fig. 2a, so that only a small part of incident energy can be guided through the air gap between neighbouring strips and then coupled through the nanoslits.

By comparison, our proposed system with the inverted $\pi$ shaped metallic grooves on the top surface of the metallic strips as illustrated in Fig. 1 can dramatically enhance the total transmission. Fig. 2b shows the field distribution of our proposed system when the geometrical parameters of the top layer are set as $W_p$ = 740 nm, $h$ = 300 nm, $B_1$ = 150 nm, $B_2$ = 180 nm, $B_3$ = 220 nm, and $B_4$ = 180 nm. It is clearly seen that the optical curtain effect is retained very well, but the field is much stronger in the output region comparing to Fig. 2a. The calculation results show that the transmission of light through the slit array is pretty high ($\eta$ = 89.1%) in Fig. 2b, which is about 3.4 times larger than that in Fig. 2a. Another evident difference between the two plots in Fig. 2b and 2a is the distribution of field on the top of the systems. Instead of highly reflection of the incident light, both the center-groove and the two side-corners strongly trap the incident energy around them in Fig. 2b.

In Fig. 3, we also plot the diffraction efficiency of reflection (thin solid), transmission (thick solid) and absorption (triangular dot) at normal incidence within the wavelength range of [0.7, 1.3] $\mu$m (the relative permittivities of silver are obtained from Ref. [36]) for the two systems in Fig. 2a and 2b, respectively. It is seen that in Fig. 3b the thick curve of the transmission efficiency reaches its maximum exactly at $\lambda$ = 1 $\mu$m. That indicates the compound groove on the top surface of the metallic nanostrip is on its resonance to effectively excite localized surface plasmon resonances as an efficient element to trap the light. In the other words, most of the incident energy will be guided into the system rather than being reflected. The calculated reflection efficiency at $\lambda$ = 1 $\mu$m is only 0.8%; see also the weak field in the incident region in Fig. 2b. However, the structure with nanostrips of the flat top surface shown in Fig. 2a is of very weak resonance and thus most of light will be reflected back toward the incident space with the reflection efficiency as high as 78%. We mention that the trapping effect benefits the extraordinary optical transmission rather than to greatly converse the energy into ohmic loss [5] because the total absorption is only around 10% at $\lambda$ = 1 $\mu$m; see the triangular dot line in Fig. 3b. We also note that because the proposed metallic nano-system with inverted $\pi$ shaped metallic grooves is on strong resonance, the absorption efficiency is much higher than that for the original structure of weak resonance as shown in Fig. 3a. In Fig. 3b, we observe another transmission peak at $\lambda = 1.1$ $\mu$m which is generated due to the particular diffraction effect of Rayleigh anomaly [39] and that is not the main content of the present work. Overall, due to the light trapping effect of the nano-corrugations, the incident energy can first be trapped in near field region and then be guided through the air gap between neighbouring inverted $\pi$ shaped metallic grooves.

In further, we replace the incident plane wave with an array of point sources beneath the inverted $\pi$ shaped metallic grooves to investigate the influence of the excitation source on the field distribution. The point sources are placed just under the grooves with the white dot of the coordinate [0, 100 nm] in Fig. 2c indicating the position at the studied period. The corresponding field distribution is plotted in Fig. 2c. Comparing Fig. 2b and 2c, we see that in the input region the field are different in distribution but both of them are of very weak intensity. Whereas, in the output region as well as in the region around the grooves, the field distributions in Fig. 2b and 2c are almost same with very strong intensity. This is because the intensity of groove generated surface plasmon resonance in the region around the inverted $\pi$ shaped metallic grooves is very strong, usually several times larger than the incident field, similar as the study in Ref. [40]. Thus the tightly trapped surface plasmon wave around the grooves are mainly determined by the geometrical parameters of the specially-shaped metallic grooves and fairly immune to the incident condition. In the input region which is a bit far from the groove surface, the groove generated surface plasmon wave gets much weaker with the intensity comparable to that of the incident wave, thus the field patterns suffer some differences as displayed in Fig. 2b and 2c.

In the following, we investigate the role of the metallic corrugations of the center-groove and the two side-corners on the effect of the extraordinary optical transmission, separately. First, we study the system with the nanostrips milled with only the center-grooves. Fig. 4a shows its transmission map when the parameters of $B_1$ and $B_2$ are tuned at normal incidence. For comparison, the corresponding map of the transmission efficiency for the proposed system with $B_3 = 220$ nm and $B_4 = 180$ nm is also plotted in Fig. 4b. Because of the existence of the side-corners, the range of $B_1$ for the proposed system in Fig. 4b is much smaller than that in Fig. 4a. In Fig. 4a and 4b, we observe that the profiles of the two maps are quite similar to each other. In detail, the transmission is at a relatively higher level when the center-groove width keeps small; this is because a narrow groove is better for trapping light than a groove with larger width [22]. In order to excite the resonance in the center-groove, choices of the groove height must allow constructive interference of $\pi$ phase length (i.e., the first order Fabry-Perot resonance) [5]. Ideally, if the phase delay at the interface of input opening is neglected, the groove height causes light propagation to experience a phase shift of $\pi/2$ and the metallic bottom of the groove gives an additional $\pi/2$ phase shift by mirror reflection. For the system with only the center-groove, the transmission reaches the maximum at $B_1 = 180$ nm, $B_2 = 140$ nm in Fig. 4a and the corresponding field distribution shown in Fig. 4c indicates that the center-groove supports the resonant mode of first order and it traps the incident light efficiently. Such light trapping phenomenon gives rise to the anti-reflection of light on the top surface of the metallic nanostrip. In Fig. 4b, the groove parameters at the maximum transmission are varied to $B_1 = 150$ nm and $B_2 = 180$ nm slightly due to the influence of the metallic side-corners while its field distributions in the center-groove as shown in Fig. 2b is very similar to that in Fig. 4c. Due to the lacking of the side-corners, the transmission map in Fig. 4a can only reach 50.7% at highest which can be clearly observed from the amplitude of field distribution in the horizontal waveguide cavity and in the output region as well as shown in Fig. 4c.

Similar study is done for investigating the role of the two side-corners. Fig. 5a and 5b show the maps of transmission efficiency with tuned $B_3$ and $B_4$ for the system with the nanostrips etched with only the two side-corners and the proposed system with $B_1 = 150$ nm and $B_2 = 180$ nm, respectively. Again, the two maps are similar to each other. In comparison with the center-groove study in Fig. 4, the transmission is optimized when the width of the side-corners is around 200 nm because the localized surface plasmon modes can be hardly excited when the corner width is too small. In Fig. 5a, the transmission efficiency reaches the maximum at $B_3 = 230$ nm and $B_4 = 220$ nm with the corresponding field distribution shown in Fig. 5c. It is seen that the side-corners get resonant with magnetic field strongly localized around their inner rectangular corners since they do not have any out-boundaries as the center-groove does. Thus, the dimension of $B_3$ times $B_4$ in Fig. 5c is a bit larger than area of $B_1$ times $B_2$ in Fig. 4c. Because of the existence of the side-corners alone, the ability of light trapping for the top surface of the metallic nanostrip in Fig. 5c is not as good as that in Fig. 2b thus the reflection of incident light is still high and the transmission efficiency of light through the nanoslits in Fig. 5c is only 42.7%. A simple calculation tells us that the addition of the transmission in Fig. 4c and 5c is approximately the value of that in Fig. 2b and by observing the field patterns, one also sees that inverted $\pi$ shaped metallic groove captures light partly by the center-groove and partly by the side-corners. Therefore, it is the combination of the center-groove and the side-corners plays the role of anti-reflecting the incident light very efficiently by supporting constructive interference to trap light into near field region. Such light anti-reflection effect is the determinative reason of the extraordinary optical transmission for the present double-layered metallic structure.

In summary, we have composed an inverted $\pi$ shaped metallic groove by corrugating one center-groove and two side-corners into the top surface of the metallic nanostrip, and then used an array of such grooves to dramatically enhance the transmission light through a metallic nanoslit array, as well as to produce an optical curtain effect in the output region. The principle of the extraordinary optical transmission caused by our designed grooves is because of the roles of light antireflection for the combination of the center-groove and the side-corners. In addition, the influence of the geometrical parameters of the center-groove and the side-corners on transmission is also investigated in detail. We mention that the optical curtain effect is retained very well even when the incident condition of light has been changed. It is expected that our study can be applied to enhance the extraction efficiency as well as control the beaming pattern of light emitting diodes.

This work is partially supported by the National Natural Science Foundation of China (11204205, 60976018, 61274056 and 60990320), the National Science Foundation (CMMI 0846771), Natural

# Figure Captions

Fig. 1. (Color online) Structure diagram of the proposed periodic metallic system comprised of inverted $\pi$ shaped metallic grooves and metallic nanoslits. The periodicity of the metallic system is $P$. The inverted $\pi$ shaped groove is formed by etching one center-groove (of width $B_1$ and height $B_2$) and two side-corners (of width $B_3$ and height $B_4$) into a metallic nanostrip (of width $W_p$ and height $h$). A metallic film with nanoslits (of width $W_s$, height $t$ and periodicity $P/2$) is placed under the grooves with a distance $d$. A plane wave with TM polarization is illuminated at normal incidence.

Fig. 2. (Color online) (a) Distribution of field $|H_y|$ for the system with un-etched metallic nanostrips. $W_p$ = 740 nm, and $h$ = 440 nm. The inset shows the relation of the transmission efficiency and the strip height for the system with un-etched metallic nanostrips. (b) Distribution of field $|H_y|$ for the proposed system with inverted $\pi$ shaped metallic grooves at normal incidence. $W_s$ = 55 nm, $t$ = 65 nm, $P$ = 1100 nm, $W_p$ = 740 nm, $h$ = 300 nm, $d$ = 140 nm, $B_1$ = 150 nm, $B_2$ = 180 nm, $B_3$ = 220 nm, and $B_4$ = 180 nm. (c) Distribution of field $|H_y|$ for the proposed system when the excitation source is an array of point sources just under the grooves and the point source within the studied period is indicated by a white dot ($x$ = 0, $z$ = 100 nm).

Fig. 3. (Color online) Diffraction efficiency of reflection (thin solid), transmission (thick solid) and absorption (triangular dot) for the system with un-etched metallic nanostrips (a) and the proposed system with inverted $\pi$ shaped metallic grooves (b) at normal incidence.

Fig. 4. (Color online) Transmission efficiency maps with tuned $B_1$ and $B_2$ for the system with the nanostrips into which only the center-groove is etched (a) and the proposed system of inverted $\pi$ shaped metallic grooves with $B_3$ = 220 nm and $B_4$ = 180 nm (b). Other parameters are set the same with those in

Fig. 2b. Different color scales are used in Fig. 3a and 3b. (c) Distribution of field $|H_y|$ for the system in Fig. 3a at $B_1$ = 180 nm and $B_2$ = 140 nm when the transmission efficiency reaches the maximum.

Fig. 5. (Color online) Transmission efficiency maps with tuned $B_3$ and $B_4$ for the system with the nanostrips into which only the two side-corners are etched (a) and the proposed system of inverted $\pi$ shaped metallic grooves with $B_1$ = 150 nm and $B_2$ = 180 nm (b). Other parameters are set the same with those in Fig. 2b. Different color scales are used in Fig. 4a and 4b. (c) Distribution of field $|H_y|$ for the system in Fig. 4a at $B_3$ = 230 nm and $B_4$ = 220 nm when the transmission efficiency reaches the maximum.

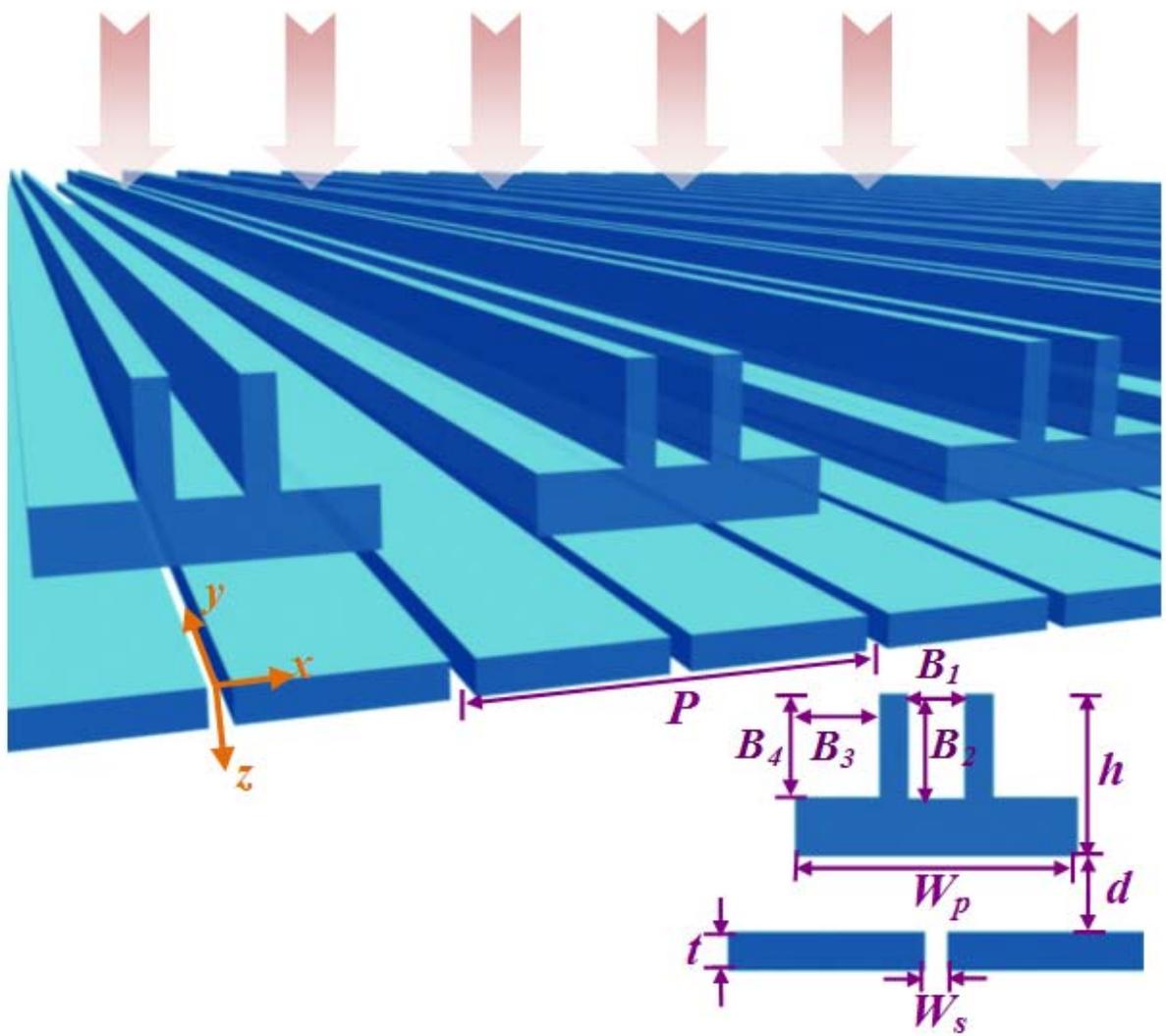

Fig. 1. Cui, Xu, et. al

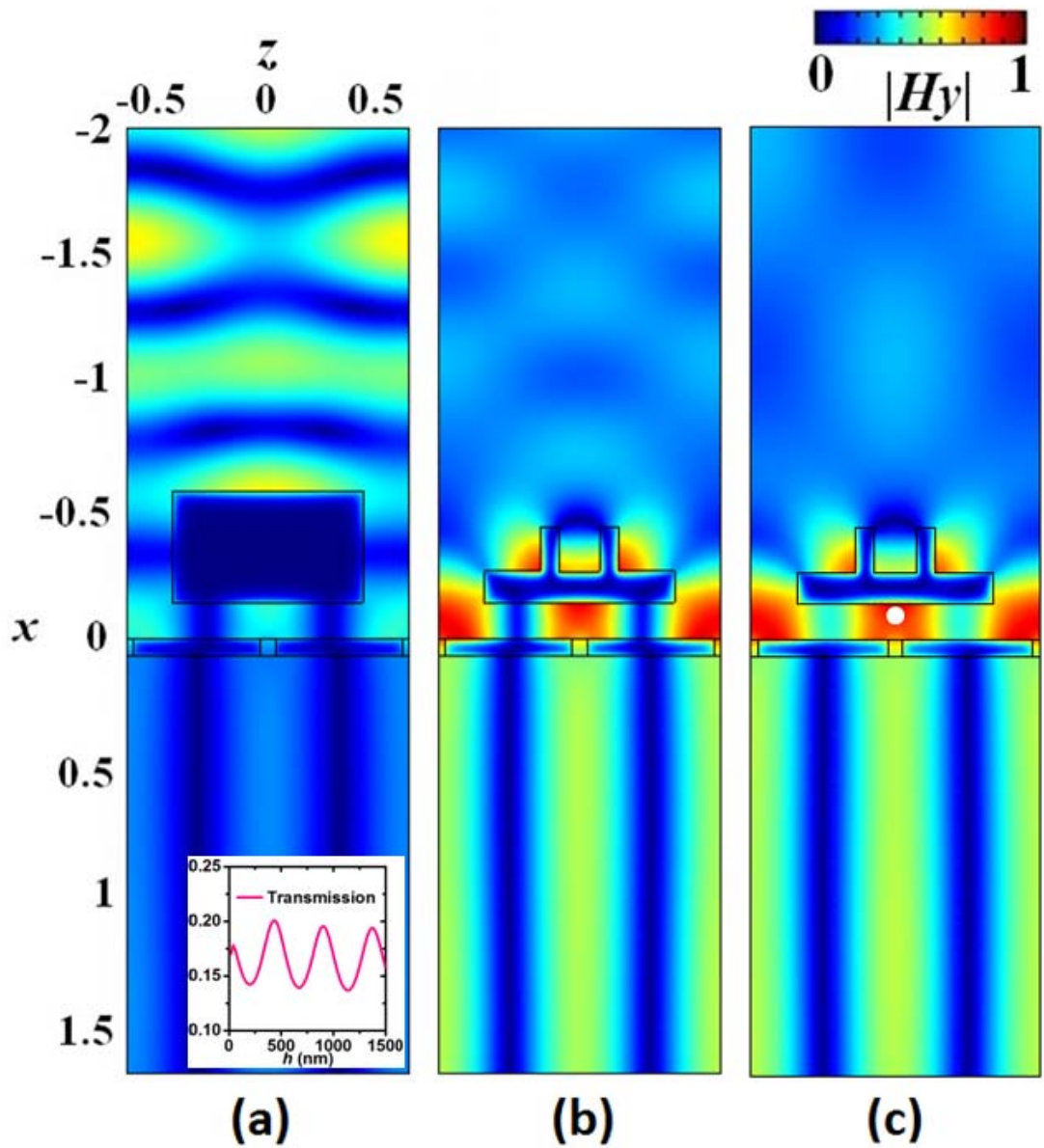

Fig. 2. Cui, Xu, et. al

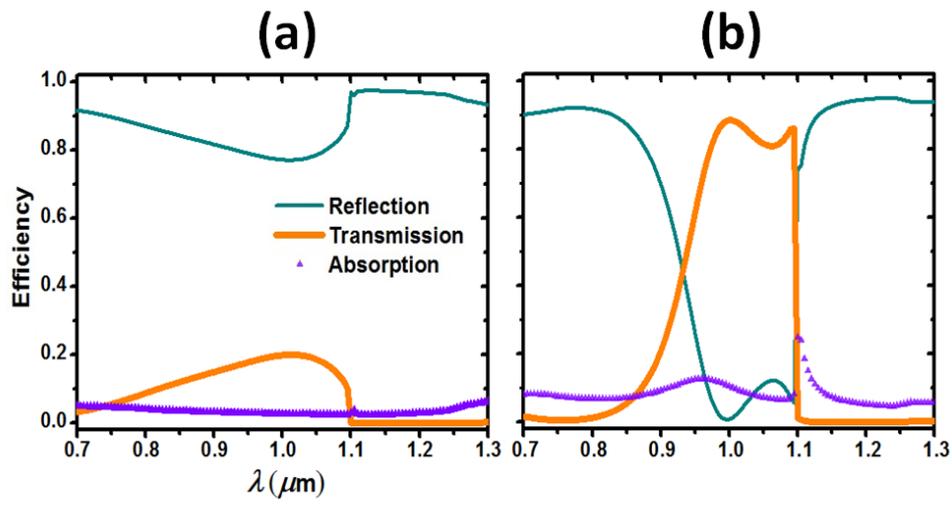

Fig. 3. Cui, Xu, et. al

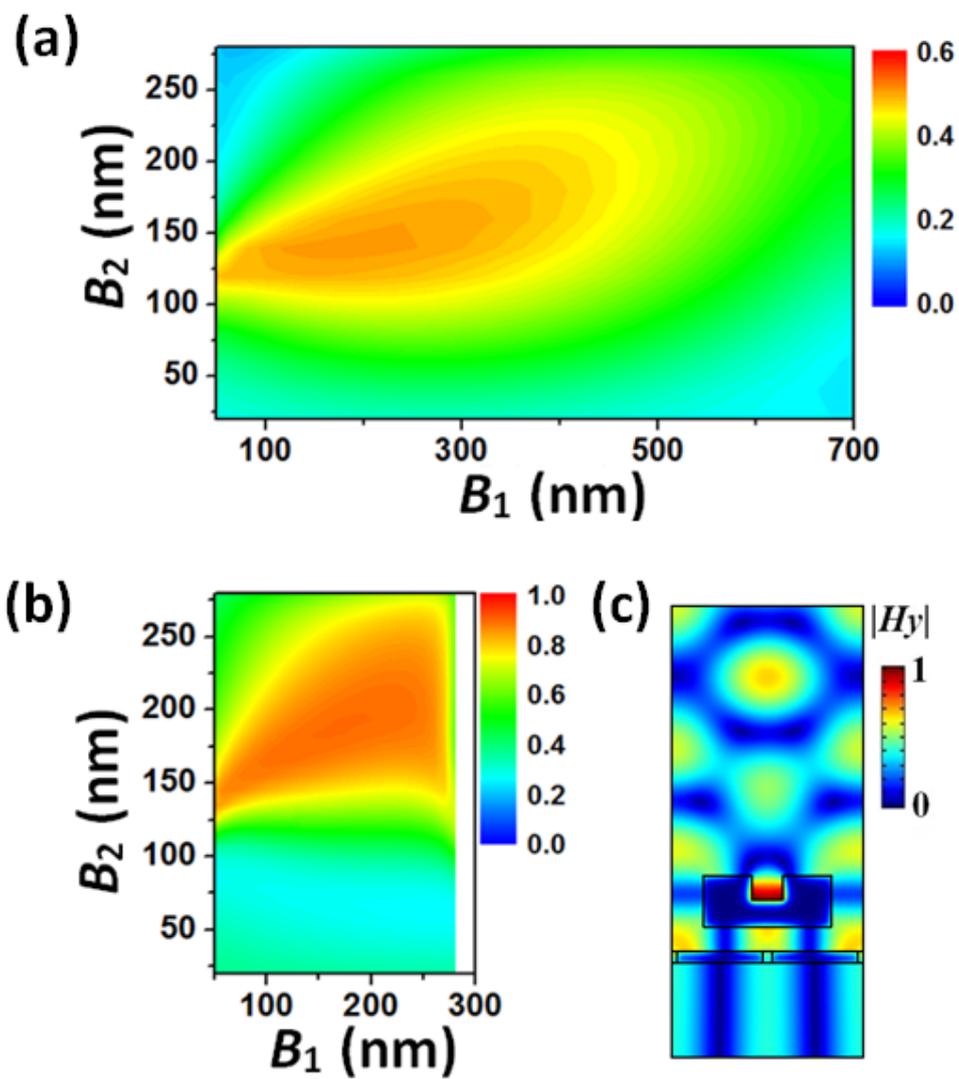

Fig. 4. Cui, Xu, et. al

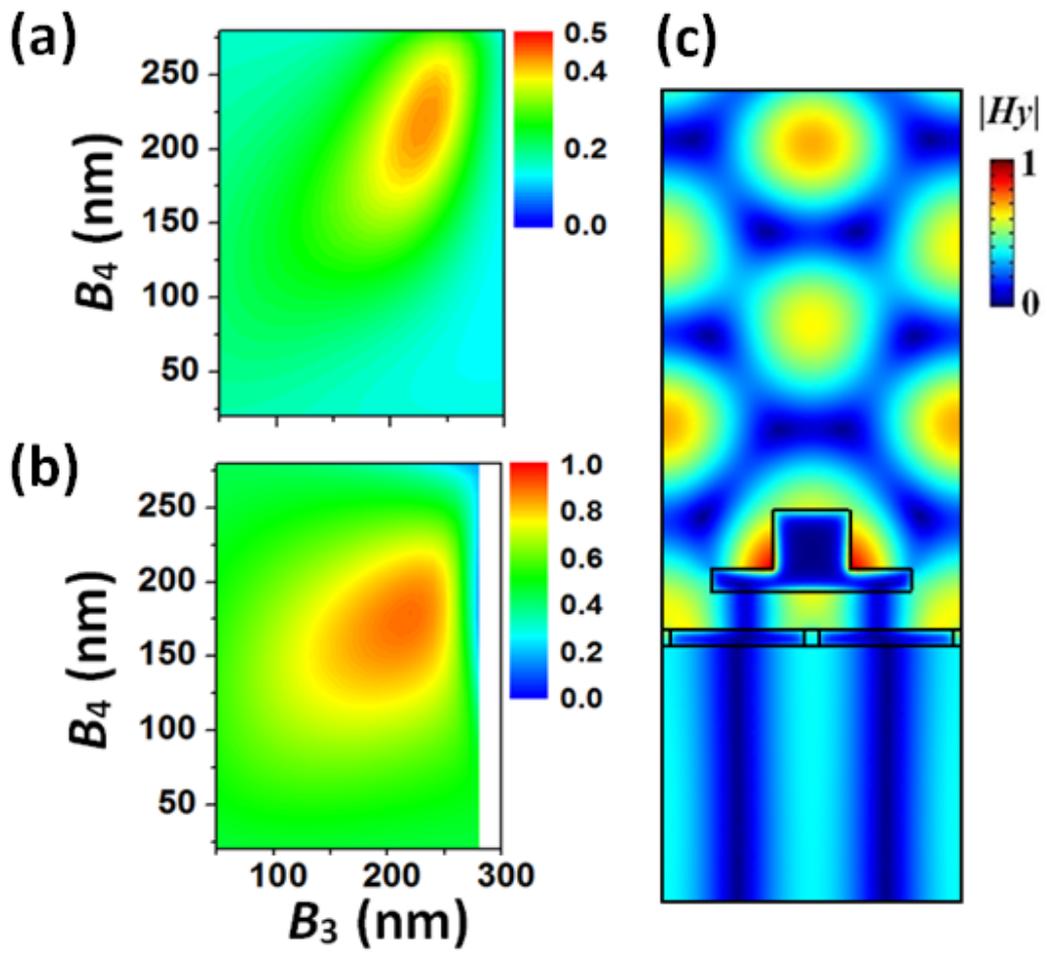

Fig. 5. Cui, Xu, et. al